\documentclass[journal=mmbox,manuscript=communication,layout=twocolumn]{achemso}

\usepackage[version=3]{mhchem} 
\usepackage{achemso,amsmath,bm,multirow,relsize}

\author{Georgios G. Vogiatzis}
\author{Doros N. Theodorou}
\email{doros@central.ntua.gr}
\affiliation[NTUA]{School of Chemical Engineering, National Technical University of Athens, 
Zografou Campus, GR-15780 Athens, Greece}

\title{Structure of polymer layers grafted to nanoparticles in silica-polystyrene nanocomposites}

\makeatletter
\let\thetitle\@title
\let\theauthor\@author
\makeatother

\makeatletter
\renewcommand\section{\@startsection{section}{1}{\z@}%
                                  {-3.5ex \@plus -1ex \@minus -.2ex}%
                                  {2.3ex \@plus.2ex}%
                                  {\normalfont\small\bfseries}}                                  
\makeatother

\makeatletter
\renewcommand\subsection{\@startsection{subsection}{1}{\z@}%
                                  {-3.5ex \@plus -1ex \@minus -.2ex}%
                                  {2.3ex \@plus.2ex}%
                                  {\normalfont\small\small\bfseries}}
\makeatother

\begin{document}
\begin{abstract}
\begin{figure} 
\begin{center}
  \includegraphics[clip,width=1.0\linewidth] {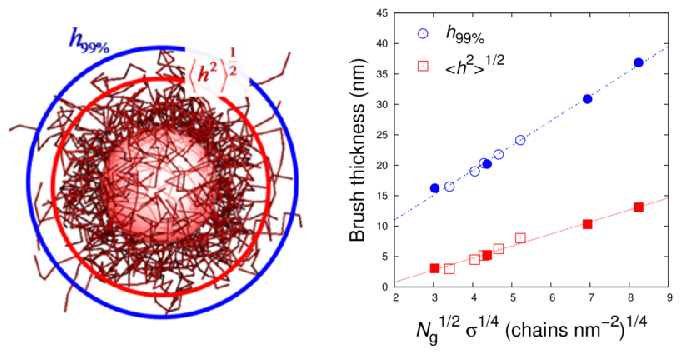}
\end{center}
\end{figure}

The structural features of polystyrene brushes grafted on spherical silica nanoparticles immersed in polystyrene
are investigated by means of a Monte Carlo methodology based on polymer mean field theory. The nanoparticle radii 
(either $8\:\text{nm}$ or $13\:\text{nm}$) are held constant, while the grafting density and the lengths of grafted 
and matrix chains are varied systematically in a series of simulations. The primary objective of this work is to 
simulate realistic nanocomposite systems of specific chemistry at experimentally accessible length scales and study 
the structure and scaling of the grafted brush. The profiles of polymer density around the particles are examined; 
based on them, the brush thickness of grafted chains is estimated and its scaling behavior is compared against 
theoretical models and experimental findings. Then, neutron scattering spectra are predicted both from single 
grafted chains and from the entire grafted corona. 
It is found that increasing both the grafting density and the grafted chain molar mass drastically alters the brush 
dimensions, affecting the wetting behavior of the polymeric brush. On the contrary, especially for particles 
dispersed in high molecular weight matrix, variation of the matrix chain length causes an almost imperceptible 
change of the density around the particle surface.
\end{abstract}

\section{Introduction}
Considerable research effort has been devoted recently to studying the structure of systems of polymers 
end-grafted onto spherical nanoparticles. 
The properties of end-constrained chains strongly differ, at high grafting densities, from those of
free chains. Tethering forces chains to depart from their free-chain, random-walk isotropic 
configuration, causing them to stretch in the direction perpendicular to the interface and to form
a brushlike structure. 
Polymer brushes find widespread industrial application as nanoparticle stabilizers. 
Recent experimental studies have shown that, \cite{Langmuir_21_6063, Macromolecules_45_4007}
most often, nanoparticles tend to aggregate into clusters or to form phase separated domains;
consequently the property improvements which might arise due to their nanoscale dimensions vanish.
The ability to disperse nanoparticles grafted with polymer brushes in a polymer matrix is of 
tremendous importance for producing polymer nanocomposites with desired mechanical and rheological 
properties. When the coverage is high enough, the nanoparticles are sterically stabilized, which results 
in good spatial dispersion. \cite{Macromolecules_29_6656,Macromolecules_45_4007}
Moreover, spherical nanoparticles uniformly grafted with macromolecules robustly self-assemble 
into a variety of anisotropic superstructures when they are dispersed in the corresponding 
homopolymer matrix. \cite{NatureMaterials_8_354}

A simpler system that is useful for understanding polymer brushes grafted on spherical nanoparticles immersed in 
melts is that of a brush grafted to a planar surface in contact with a melt of chemically identical chains.  Important 
molecular parameters for this system are the Kuhn segment length of the chains, $b$, the lengths (in Kuhn segments)
of the grafted, $N_{\rm g}$ and free, $N_{\rm f}$, chains, and the surface grafting density (chains per unit 
area), $\sigma$. 
The case of planar polymer brushes exposed to low molecular weight solvent was studied theoretically
by de Gennes \cite{Macromolecules_13_1069} and Alexander \cite{JPhysFrance_38_983}.
These authors used a scaling approach in which a constant density was assumed throughout the brush: all the brush
chains were assumed to be equally stretched to a distance from the substrate equal to the thickness of the brush.
Aubouy et al. \cite{Macromolecules_28_2979} extracted the phase diagram of a planar brush exposed to 
a high molecular weight chemically identical matrix.
Their scaling analysis is based on the assumption of a steplike concentration profile 
and on imposing the condition that all chain ends lie at the same distance from the planar surface. 
Five regions with different scaling laws for the height, $h$, of the brush were identified. 
For low enough grafting densities, $\sigma < N_\text{g}^{-1}a^{-2}$ (with
$a$ being the monomer size) and short free chains, $N_\text{f} < N_\text{g}^{1/2}$, the brush behaves as
a swollen mushroom, with $h \sim N_{\rm g}^{3/5}$. 
By keeping the grafting density below $N_{\rm g}^{-1}a^{-2}$ and increasing $N_{\rm f}$, so that
$N_{\rm f}>N_{\rm g}^{1/2}$, the brush becomes ideal with $h \sim N_{\rm g}^{1/2}$. 
For intermediate grafting densities, $N_{\rm g}^{-1} < \sigma < N_{\rm g}^{-1/2}$, high 
molecular weight free chains, $N_{\rm f}>N_{\rm g}^{1/2}$, can penetrate the brush, thus ideally 
wetting it and leading to $h \sim N_{\rm g}^{1/2}$. Increasing the grafting density while keeping 
$N_{\rm f}<N_{\rm g}^{1/2}$ forces the chains to stretch, thus leading to a brush height scaling 
as $h \sim N_{\rm g}$. 

A numerical Self Consistent Field (SCF) calculation has also been reported \cite{Macromolecules_20_1692},
in which the density profile is no longer assumed to be a step profile and where the end points
of the chains are distributed throughout the brush.
Analytical equations based on a similar model were developed by Milner et al. \cite{Macromolecules_21_2610}
and by Zhulina et al. \cite{Macromolecules_28_1491}
In the wetting state, the grafted and free chains are intermixed, along the full extent of the brush.
If the matrix and the grafted chains demix, then the corona collapses and the brush is dewetted.
A detailed study of moderately stretched planar brushes exposed to moderately long melt chains 
by Ferreira et al. \cite{Macromolecules_31_3994} found that the domain where attraction exists between 
two grafted layers in a melt, and where partial wetting is thus expected, scales as 
$\sigma \sqrt{N_\text{g}} > (N_\text{g}/N_\text{f})^2$. 
This scaling law indicates that flat surfaces grafted with sparse polymer brushes in a long chain polymer 
melt could exhibit entropic attraction, provided that the molecular weight of the matrix is large enough.
Experimental \cite{Macromolecules_29_2150}, SCF \cite{Macromolecules_31_3994} and Molecular Dynamics (MD) \cite{JChemPhys_105_5532} works on planar polymer brushes, in a chemically identical matrix, 
have shown that the matrix wets the polymer brush only when the melt chains are shorter than the chains of the brush.
It is experimentally observed \cite{Langmuir_18_8871} that ``autophobic dewetting'' \cite{JChemPhys_115_2794} occurs when 
the brush and the matrix share the same length. Longer melt chains spontaneously dewet the brush, because the
gain in mixing entropy cannot overcome the conformational entropy loss associated with the matrix chains penetrating 
the brush.\cite{Macromolecules_29_6656}

Similar ideas \cite{Macromolecules_26_7214} can be used to understand the configurations of polymer brushes
grafted to spherical nanoparticles dispersed in a polymer melt. Here the radius of the nanoparticle, 
$R_{\rm n}$, enters as an additional parameter.  
Long polymers grafted to a surface at fixed grafting density, $\sigma$, are strongly perturbed from their ideal
random-walk conformation.\cite{Macromolecules_24_693}
Planar geometry scaling (infinite curvature) is inadequate to explain the case when the particle size is 
reduced to a level comparable with the size of the brushes.
The SCF theory has been applied to convex (cylindrical and spherical) surfaces by Ball et al. \cite{Macromolecules_24_693}
For the cylindrical case, under melt conditions, it was found that the free ends of grafted chains are excluded from a zone
near the grafting surface. The thickness of this dead zone varies between zero for a flat surface to a finite fraction 
of the  brush height, $h$, in the limit of strong curvature, when $R_\text{n}/h$ is of order unity, with $R_\text{n}$ 
being the radius of curvature of the surface. 
Borukhov and Leibler \cite{Macromolecules_35_5171} presented a phase diagram for brushes 
grafted to spherical particles, in which the five regions of the work of Aubouy et al. can still be  
located, but they also provide the scaling of the exclusion zone, where matrix chains are not present.
Such exclusion zones have been observed in a special limiting case of grafted polymers, namely, star shaped polymers.
Daoud and Cotton \cite{JPhys_43_531} showed that, in the poor-solvent limit, the free ends of the chains are pushed outward, 
because of high densities near the center of the star. The Daoud-Cotton model assumes that all chain ends are a uniform
distance away, while the Wijmans-Zhulina model \cite{Macromolecules_26_7214} has a well-defined exclusion zone.
For $\Theta$ solvents, in the limit of large curvature (small particle radius, $R_\text{n}$), the 
segment density profile, $\phi(r)$, decreases with the radius as \cite{Macromolecules_26_7214}
$\phi(r) \propto \sigma^{1/2} \left(R_\text{n}/r\right)$.
It must be noted that $\phi$ is not linear in $\sigma$ because the brush height depends on $\sigma$.
In the limit of small curvature (large $R_\text{n}$), a distribution of chain ends must be accounted
for,\cite{Macromolecules_21_2610} leading the segment density profile to a parabolic form: \cite{Macromolecules_26_7214}
$\phi(r') = \frac{3\sigma N_\text{g} b^3}{h_0}\left(\frac{h}{h_0}\right)^2\left(1-\left(\frac{r'}{h}\right)^2\right)$ 
where $b$ is the statistical segment length, $r'= r-R_n$ is the radial distance from the particle surface, $h_0$ is 
the effective brush height for a flat surface and $h$ is the brush thickness.
For large nanoparticles, the above form asymptotically recovers the planar result ($h \to h_0$).
In the case of intermediate particle radii, a combination of large and small curvature behaviors is anticipated:
\cite{Macromolecules_26_7214} the segment density profile exhibits large curvature behavior near the surface of
the particle, followed by a small curvature behavior away from it.
Finally, following Daoud and Cotton, the brush height is expected to scale as $h \propto \sigma^{1/4} N_\text{g}^{1/2}$.

Hasegawa et al.\cite{Macromolecules_29_6656} used rheological measurements and SCF calculations to show that 
particles are dispersed optimally when chains from the melt interpenetrate, or wet, a grafted polymer brush (``complete wetting''). 
This occurs at a critical grafting density, which coincides with the formation of a stretched polymer brush on the 
particle surface.\cite{Macromolecules_13_1069, JPhysFrance_38_983}
Grafting just below this critical density produces aggregates of particles due to attractive van der Waals forces between them.
The results of these authors suggest that mushrooms of nonoverlapping grafted polymer chains have no ability to stabilize 
the particles against aggregation (``allophobic dewetting''). 
Grafting just above this critical density also results in suboptimal dispersions, the aggregation of the particles 
now being induced by an attraction between the grafted brushes.
For high curvature (small radius) nanoparticles, the polymer brush chains can explore more space, resulting in less entropic
loss to penetrate the brush, reducing the tendency for autophobic dewetting. 

Detailed MD simulations by Peters et al. \cite{Langmuir_28_17443} have been used to study the effect of passivating
ligands of varying molar mass grafted to a 5-nm-diameter amorphous silica nanoparticle and placed in various 
alkane solvents. 
Their study focused on the average height and density profiles for methyl-terminated alkoxysilane 
ligands exposed to either explicitly modeled, short-chain hydrocarbon solvents or implicit good and poor solvents.
Coating brushes equilibrated in explicit solvents were more compact in longer-chain solvents because of 
autophobic dewetting. The response of coatings in branched solvent was not significantly different from that in the 
linear solvent of equal mass. Significant interpenetration of the solvent chains with the brush coating was observed 
only for the shortest grafted chains.
An implicit poor solvent captured the effect of the longest chain length solvent at lower temperatures, while 
an implicit good solvent produced coating structures that were far more extended than those found in any of the 
explicit solvents tested and showed little dependence on temperature. 

A major challenge in simulating nanocomposite materials with chemically realistic models
is that both the length and the time scales cannot 
be adequately treated by means of atomistic simulations because of their extensive computational requirements. 
This is why a variety of mesoscopic techniques have been developed for these particular systems.
Kalb et al.\cite{SoftMatter_7_1418} and later Meng et al.\cite{SoftMatter_8_5002} have performed Molecular Dynamics
simulations where both brush and melt chains were represented using the coarse-grained bead-spring 
model of Kremer and Grest.\cite{PhysRevLett_61_566}
One of their conclusions was that the conformational transition between a stretched ``wet'' brush and
a collapsed ``dry'' brush, which has been attributed to the autophobic dewetting of the brush by the 
melt, was not readily apparent in their brush density profiles. The calculated Potential of 
Mean Force (PMF) was found to go from purely repulsive to attractive by increasing the molar mass
of the matrix with a well depth on the order of $k_\text{B}T$. These results were purely entropic
in origin and arose from the competition between brush-brush repulsion and an attractive inter-nanoparticle
interaction caused by matrix depletion from the inter-nanoparticle zone.
Integral equation approaches, such as the polymer reference interaction site model (PRISM),
have provided important insight and detailed information about inter-particle potentials and spatial 
correlations between the locations of the particles. As Schweizer and co-workers have shown,
\cite{Macromolecules_39_5133, Macromolecules_41_9430} there is a competition between tether-mediated 
microphase separation and matrix-induced macrophase separation for nanoparticles dispersed in a polymeric matrix.

Significant progress has been made in applying the SCF theory to nanoparticle-filled polymer matrices in 
the dilute and semi-dilute region.\cite{Langmuir_22_969, PhysRevE_78_051804, SoftMatter_8_4185,EurPolymJ_49_613}
An attempt to overcome the restrictions posed by the saddle-point approximation was made 
by Sides et al.\cite{PhysRevLett_96_250601} 
The coordinates of all particles in the system were explicitly retained as degrees of freedom. 
It was crucial to update simultaneously the coordinates and the chemical potential field variables.
Laradji et al. \cite{PhysRevE_49_3199} introduced the idea of evolving the explicit conformations of the 
molecules by standard Metropolis Monte Carlo, while obeying an Edwards mean-field Hamiltonian.
The paper of Laradji et al. was pioneering in introducing the concept of conducting particle-based simulations 
with interactions described via collective variables and also two ways of calculating these collective variables:  
grid-based and continuum weighting-function-based. A clearer discussion of this point is found in studies of brushes in 
poor solvent by Soga et al.\cite{EurophysLett_29_531}
Later, Daoulas and M\"uller\cite{JChemPhys_125_184904} and Detcheverry et al.\cite{Macromolecules_41_4989} 
have extensively used this methodology to study a wide range of polymeric systems.
This technique has been very successful in describing the self-assembly and the ordering of block 
copolymers on patterned substrates encountered in processes such as nanolithography.\cite{Macromolecules_43_2334}
Harton and Kumar\cite{JPolymSciPartBPolymPhys_46_351} have developed a Flory theory, coupled with the self-consistent 
field model of Scheutjens and Fleer,
to probe the miscibility of homopolymer-grafted nanoparticles with homopolymer matrices.
Their main conclusions were that miscibility is improved with decreasing particle radius and increasing brush chain length.
They have also found that greater swelling of the brush by matrix chains is required for dispersion
of larger brush-coated nanoparticles.

Striolo and Egorov \cite{JChemPhys_126_014902} have employed both MC simulations and Density Functional
Theory (DFT) calculations to investigate the interactions between spherical colloidal brushes both in 
vacuum (implicit solvent) and in 
explicit solutions of nonadsorbing polymers. These authors found that interactions between hard sphere particles
grafted with hard-sphere chains are always repulsive in implicit solvent. The range and steepness of the 
repulsive interaction is sensitive to the grafting density and length of the grafted chains.
When the brushes are immersed in an explicit solvent of hard-sphere chains, a weak mid-range attraction arises,
provided the length of the free chains exceeds that of the grafted chains. This attraction may be due the 
depletion of free chains from the region between the brushes.
A comparison between the two methods indicated that DFT is in semiquantitative agreement with MC results, but 
required a small fraction of the computer time needed by MC.

Until recently, the experimental verification of theoretical and simulation predictions was mostly 
limited to global information concerning the brush, such as its average height, but not its 
profile.\cite{Macromolecules_43_1564} Recently Chevigny et al.\cite{Macromolecules_43_4833} used 
a combination of X-ray and Small Angle Neutron Scattering (SANS) techniques to measure the 
conformation of chains in polystyrene polymer brushes grafted to silica nanoparticles with an 
average radius of $13\:\text{nm}$ dispersed in polystyrene matrix.
They found that, if the molecular weight of the melt chains becomes large enough, the polymer 
brushes are compressed by a factor of two in thickness compared to their stretched conformation 
in solution. 
Also, polymer brushes exposed to a high molecular weight matrix are slightly compressed 
in comparison to brushes exposed to a low molecular weight matrix environment. 
This observation implies a wet to dry conformational transition. The low molar mass free chains can penetrate into the 
grafted brush and swell it (``wet'' brush). Conversely, when grafted and free chain molar masses are comparable,
free entities are expelled from the corona (``dry'' brush). 
Later, they examined the dispersion of these grafted particles in melts of different 
molar masses, $M_\text{f}$.\cite{Macromolecules_44_122}
They showed that for $M_\text{g}/M_\text{f}<0.24$, the nanoparticles formed a series of compact aggregates,
whereas for $M_\text{g}/M_\text{f}>0.24$, the nanoparticles were dispersed within the polymer host.

The present work examines the structure of atactic polystyrene brushes grafted on silica nanoparticles, 
dispersed at the limit of infinite dilution,in polymer matrices of the same chemical constitution (atactic polystyrene).
Particular attention is given to capturing the chain conformation of both matrix and grafted chains. 
The analysis is based on a coarse-grained Monte Carlo simulation method which has already been applied to a 
complex three-dimensional polymer-polymer nanocomposite system.\cite{EurPolymJ_47_699}
Although the level of description is coase-grained (e.g., employing freely-jointed chains to represent the matrix), 
the approach developed aims at predicting the behavior of a nanocomposite with specific chemistry quantitatively, 
in contrast to previous coarse-grained simulations.
A main characteristic of the method is that it treats polymer-polymer and polymer-particle interactions in a 
different manner: the former are accounted for through a suitable functional of the local polymer density, while the 
latter are described directly by an explicit interaction potential.
The term ``fast off-lattice Monte Carlo (FOMC) simulations'' accounts for this fact.\cite{JChemPhys_130_104903} 
The basic idea of FOMC simulations
is to use ``soft'' potentials based on mean-field considerations that allow
particle overlapping in continuum MC simulations. For multichain systems, this gives much faster chain relaxation
and better sampling of the configurational space than conventional molecular simulations using ``hard'' excluded-volume
interactions that prevent particle overlapping.
The main advantage of this approach over  SCF is that can directly sample polymer chain conformations, 
which is a formidable task for pure SCF calculations.
Our efforts will focus on the wetting/dewetting transition of the grafted polymer brush, which is of great
importance, since it is directly related to the phase separation/miscibility of the nanocomposite systems.
Moreover, neutron scattering from these systems will be predicted from a discrete Fourier transform of the 
real space coordinates of the whole system. To the best of our knowledge, this is the first time an 
analysis of this kind is undertaken for realistic nanocomposite systems matching experimentally studied ones.

\section{Model and simulation methodology}
\subsection{Model}
We provide here some details on the FOMC model, which was introduced in ref \citenum{EurPolymJ_47_699}.
The configurational partition function on which the Monte Carlo sampling is based is expressed as a 
functional integral over the paths of all chains and as an integral over all nanoparticle positions 
and orientations. For spherical nanoparticles, orientations play a role only when grafted chains are present. 
\begin{align}
\mathcal{Z}  = & \frac{1}{n_\text{f}!} \mathlarger{\int} 
\prod_{i=1}^{n_\text{f}} \left(\mathcal{D}\left[\mathbf{r}_i(s) \right] 
\mathcal{P}_0\left[\mathbf{r}_i (s) \right]\right) \nonumber \\
& \prod_{k=1}^{n_\text{p}} \left(d^3 \mathbf{R}_k d^2 \pmb{\Omega} _k \right) \:
\prod_{j_k =1}^{n_{g_k}} \left(\mathcal{D}\left[\mathbf{r}_{j_k}(s)\right] \mathcal{P}_0 
\left[\mathbf{r}_{j_k} (s);\mathbf{R}_k,\pmb{\Omega}_k \right]\right) \: \nonumber \\
&\times 
\exp \left(- \beta \mathcal{H}_{\rm p} \left[\left\{\mathbf{R}_k\right\}, \left\{\mathbf{r}_i(s)\right\}, 
\left\{\mathbf{r}_{j_k}(s) \right\}\right] -\beta \mathcal{H}_{\rm nb} \left[\widehat{\rho} (\mathbf{r})\right] \right)
\end{align}
Here $n_\text{p}$ is the total number of nanoparticles, $n_\text{f}$ is the total number of free chains, $n_{\text{g},k}$ is the number
of chains grafted to nanoparticle $k$, and $\beta=1/\left(k_\text{B}T\right)$. $\mathbf{R}_k$ and $\pmb{\Omega}_k$ stand for 
the center of mass position and orientation (Euler angles) of nanoparticle $k$, respectively.
Each one of the $n_{\text{g},k}$ chains that are grafted to nanoparticle $k$ is rigidly affixed to its surface by one of 
its ends.
The statistical weights $\mathcal{P}_0\left[\mathbf{r}_i(s)\right]$ and 
$\mathcal{P}_0 \left[\mathbf{r}_{j_k}(s); \mathbf{R}_k,\pmb{\Omega}_k \right]$ incorporate bonded interactions along each one of the 
chains; in the implementation employed here, they correspond to freely jointed chains. 

The effective Hamiltonian $\mathcal{H}_\text{nb}\left[\widehat{\rho}\left(\mathbf{r} \right) \right]$ represents 
nonbonded interactions among chains, either free or grafted. 
An approximation based on the mean-field averaging out of small scale fluctuations is applied to this latter
contribution. 
Instead of expressing the nonbonded interaction energy as a sum of pairwise interactions 
between segments, as is done in atomistic and coarse-grained simulations, we assume that it is as a functional of the 
three-dimensional density distribution of polymer segments, $\widehat{\rho} \left(\mathbf{r} \right)$. 
The density distribution is calculated by partitioning the system into cells using a cubic grid of spacing $\Delta L$.
We then assign each polymer segment to the center of the cell to which it belongs. The contribution from terminal 
segments is assumed to be half that of the interior segments in this assignment. The volume of each cell is taken as 
$\left(\Delta L \right)^3$ minus the volume of any sections of nanoparticles that find themselves in the cell; it is 
computed via a fast analytic algorithm \cite{MolPhys_72_1313}. 
Following early work by Helfand and Tagami,\cite{JChemPhys_56_3592}
a functional of local density is derived heuristically guided by the macroscopic thermodynamic behavior of the system.
The nonbonded effective Hamiltonian, punishes departures of the local density in each cell from the mean segment 
density $\rho_0$ in the melt under the temperature and pressure conditions of interest:
\begin{equation}
\mathcal{H}_\text{nb}\left[\widehat{\rho} \left(\mathbf{r} \right) \right] = \frac{\kappa_0 \left(\Delta L \right)^3}{2\rho_0}
\sum_{m=1}^{N_\text{cells}}\left(\widehat{\rho}_m - \rho_0 \right)^2
\label{eq_field_hamiltonian}
\end{equation}
The prefactor $\kappa_0$ is related to the isothermal compressibility $\kappa_{\rm T}$ of the melt via 
$\kappa_0 = 1/(k_\text{B} T \kappa_{\rm T} \rho_0)$.

The effective Hamiltonian $\mathcal{H}_\text{p}\left[\left\{\mathbf{R}_k\right\}, 
\left\{\mathbf{r}_i(s)\right\}, \left\{\mathbf{r}_{j_k}(s)\right\} \right]$ encompasses, in general, 
nanoparticle-nanoparticle and polymer-nanoparticle interactions.  In the infinitely dilute nanoparticle case considered 
here, it embodies interactions between segments of the free and grafted chains with the nanoparticles:
\begin{align}
& \mathcal{H}_\text{p}\left[\left\{\mathbf{R}_k\right\}, \left\{\mathbf{r}_i(s)\right\}, 
\left\{\mathbf{r}_{j_k}(s)\right\} \right] = \nonumber \\
& \sum_{k=1}^{n_{\rm p}} 
\left\{  
\sum_{i=1}^{n_{\rm f}} \sum_{s=0}^{N_{\rm f}} \mathcal{V}_{\rm ps} \left(\mathbf{r}_i(s) - \mathbf{R}_k  \right)
\right .
\nonumber \\
& \left. + \sum_{l=1}^{n_{\rm p}} \sum_{j_l=1}^{n_{{\rm g}_l}} \sum_{s=0}^{N_{\rm g}} \mathcal{V}_{\rm ps}
\left( \mathbf{r}_{j_l}(s) - \mathbf{R}_k \right)
 \right\}
\end{align}
It is calculated from the center of mass positions and radii of the nanoparticles and polymer segments 
and from the densities of the atoms constituting each nanoparticle and each polymer segment, as a sum
of Hamaker integrated potentials, $\mathcal{V}_{\rm ps}$.
The summation extends over all free and grafted segments of the system, excluding only the interaction 
between the segments which are directly connected to a particle and the particle itself. 

The interaction of a polymeric segment and a nanoparticle can 
be calculated using eqs 11-13 of ref \citenum{EurPolymJ_47_699}, which are restated in the Supporting
Information to the present paper, with suitable Hamaker constants for the polymer and the particle. 
The Hamaker constant governing the interaction between two spherical bodies can be calculated 
as \cite{PhysRevE_67_041710} 
$A = 4 \pi^2 \epsilon_{\rm LJ} \left(\rho_{\rm LJ} \sigma_{\rm LJ}^3 \right)^2$,
where $\epsilon_{\rm LJ}$, $\sigma_{\rm LJ}$ are the atomistic Lennard-Jones interaction parameters,
and $\rho_{\rm LJ}$ the density of atomistic interaction sites in the macroscopic body.
The silica particle contributes only the interaction of its oxygens with the polymeric matrix 
($\epsilon_{\rm O}$,$\sigma_{\rm O}$ for silica \cite{JChemPhys_104_6319}), while atomistic Lennard-Jones 
parameters for \ce{CH}(aromatic), \ce{CH2}(aliphatic) and \ce{CH3}(aliphatic) sites of polystyrene are used.
\cite{Macromolecules_40_3876}
For every kind of interaction site of polystyrene, the corresponding Hamaker constant is calculated
by assuming $\rho_{\rm LJ}$ to be the density of the corresponding kind of sites in the bulk polymer.
Electrostatic interactions are not taken into account either for silica or for polystyrene.
Hamaker constants, obtained here by integration of the Lennard-Jones potentials used
in atomistic simulations of the system, are within experimental range for polystyrene 
\cite{ColloidPolymSci_256_343,JColloidInterfaceSci_115_463} and silica.\cite{AdvColloidInterfaceSci_70_125}
The corresponding values are provided in the ``Simulation Details'' subsection.

The reference level of the employed Hamiltonian ( \ref{eq_field_hamiltonian}) for the homopolymer
melt is that of a melt with uniform density profile. In this case, the effective interaction energy 
between two polymeric beads is taken as zero.
The effective Hamaker constant of polymer-nanoparticle interaction should be such that, if the volume
of the nanoparticle were occupied by bulk homopolymer, then the energy of the system would be zero. 
The insertion of a nanoparticle in a uniformly-distributed melt can be thermodynamically accomplished 
in two steps: a spherical volume of polymer, whose net interaction with the rest of the polymer matrix 
depends on $A_{\rm PS}$, is removed from the melt and a nanoparticle with equal volume is placed in 
its position, introducing a new net interaction depending on $\sqrt{A_\text{PS} \: A_\text{\ce{SiO2}}}$ 
with the remaining bulk polymer.\cite{Hiemenz} 
Thus, the Hamaker constant of the effective interaction can be calculated as: 
\begin{equation}
A_\text{PS-\ce{SiO2}}^{\rm eff} = \sqrt{A_\text{PS} \: A_\text{\ce{SiO2}}} - A_\text{PS}\;,
\end{equation}
and used for the nanoparticle - polymeric segment interaction, $\mathcal{V}_{\rm ps}$, following the expressions
presented in the Supporting Information to the present paper.

\subsection{Initial Configuration} 
To start the FOMC simulation, an initial configuration is generated by placing the nanoparticles at randomly selected 
positions, so that they do not overlap, and then building the polymeric chains around them. The first chains to be 
built are the grafted ones. Initially, the grafting points are randomly selected on the surface of the particle. Then, 
these points are allowed to move freely on the surface of the particle (via a MD-like procedure), under the influence 
of a repulsive pseudo-potential between them. This step is undertaken in order to mimic the more or less uniform 
distribution of initiators and linkers connecting grafted chains to the nanoparticle surface, observed experimentally. 
Both grafted and free chains are grown as random walks in a segment-by-segment fashion. When the addition of a segment 
fails (due to the constrained environment imposed by the particle), a geometric algorithm is used in order to drive 
the insertion of the segment towards higher free-volume regions. The initial configurations created by this method 
exhibit large local density fluctuations. These fluctuations increase with increasing chain length. 
A zero temperature Monte Carlo optimization procedure takes place in order to reduce the density fluctuations.
During this stage, all moves leading to a more uniform density profile, thus decreasing the density fluctuations,
are accepted. In the opposite case, they are rejected.\cite{JChemPhys_119_12718}

\subsection{Equilibration} 
Equilibration during a FOMC run is achieved through MC based on the probability density associated with the partition 
function $\mathcal{Z}$ stated above. The fact that chain-chain interactions are accounted for only approximately 
through $\mathcal{H}_\text{nb}$, allows bold rearrangements of the chain conformations to be attempted with significant 
probability of success. Rigid displacements and rotations of the chains, mirroring transformations through planes 
passing through the chain centers of mass, and exchanges of entire chains keeping their center of mass positions fixed, 
are used.\cite{JChemPhys_119_12718} In addition, the internal conformations of chains are sampled using flips of internal 
segments,\cite{EurPolymJ_47_699} end segment rotations,\cite{EurPolymJ_47_699} reptation\cite{JChemPhys_63_4592} 
and pivot\cite{MolPhys_17_57} moves. 
Grafted chains equilibrate only through flip, end rotation and pivot moves, which keep their grafted end constrained.
An additional Monte Carlo move entails attempting random translations of the origin of the grid used for the estimation 
of local densities,  within a cube of edge length $\Delta L$.  Following each such translation the energy of the system 
is recalculated and the move is accepted or rejected according to the standard Metropolis selection criterion. This grid 
translation move has been included in order to avoid artifacts due to spatial discretization. The exact mixture of 
moves used in the simulations and several measures of equilibration are reported in the Supporting Information to 
the present paper.

\subsection{Simulation Details} 
The systems considered were dilute dispersions of silica nanoparticles with atactic polystyrene chains grafted 
on their surface, in atactic polystyrene melts. Molecular parameters needed for the FOMC calculation, as explained above, 
are summarized in Table \ref{tab:ftimc_parameters}. The radius of the nanoparticles, $R_\text{n}$, was either $8\:\text{nm}$ 
or $13\:\text{nm}$. Both grafted and free chains of polystyrene were strictly monodisperse. The 
molecular weight of the free chains, $M_\text{f}$, and of the grafted chains, $M_\text{g}$, was varied, and the 
corresponding numbers of Kuhn segments, $N_\text{f}$ and $N_\text{g}$, were calculated based on the molar mass of a 
Kuhn segment, listed in Table \ref{tab:ftimc_parameters}. The molecular characteristics of the systems studied, i.e., the 
degree of polymerization of the chains, the radius and grafting density of the silica nanoparticles, have been selected 
in such a way as to match systems which have been or can be studied experimentally 
\cite{SoftMatter_5_7741,Macromolecules_43_4833,MeyerThesis}. All simulations were carried out in the canonical statistical ensemble. 
The temperature of the system was $ T=500 \text{K}$. The simulation box was cubic with varying edge length from 
$80\:\text{nm}$ to $200\:\text{nm}$. Within such a box a single nanoparticle was placed, thus leading to a volume 
fraction of nanoparticles, $\phi$, less than 1\%. It should be noted that, for all chain lengths, the edge length is 
at least an order of magnitude larger than $R_{\text{g},0}$, with $R_{\text{g},0}$ being the unperturbed radius of 
gyration of a chain, so as to avoid finite size effects ($R_{\text{g},0}\simeq 8.7 \:\text{nm}$ for 100 kg/mol PS). 
The grafting density, $\sigma$ varied from $0.2 \:\text{nm}^{-2}$ to $1.1 \:\text{nm}^{-2}$.  It was manipulated by
defining the number of grafted chains, $n_\text{g}$, on the nanoparticle surface. 
Most experimental work up to now has been concentrated on low grafting densities (around $0.2\;{\rm nm}^{-2}$).
\cite{NatureMaterials_8_354,SoftMatter_5_7741} However, silica particles with higher grafting densities (around 
$1.0\;{\rm nm}^{-2}$) coated with asymmetric block copolymers have also been synthesized.\cite{Macromolecules_43_8029}
The number of free polymer chains in the system, $n_\text{f}$, was varied according to the chain length so that the 
mean density of polymer in the accessible volume of the model system was $0.953 \:\text{g/cm}^3$, matching the 
experimentally measured value \cite{TransFaradaySoc_67_2251}. The Kuhn segment length, $b$, is $1.83\:\text{nm}$, 
which corresponds to seven styrene monomers. 

\begin{table}
   \caption{FOMC simulation parameters}
   \label{tab:ftimc_parameters}
   \begin{tabular}{cc}
      \hline
      Parameter & Value \\ \hline 
      Temperature, $T$                    & $500\:\text{K}$ \\
      Mean polymer density, $\rho_0$      & $0.953\:\text{g/cm}^3$ \\
      Amorphous silica density, $\rho_\text{\ce{SiO2}}$ & $2.19 \:\text{g/cm}^3$ \\
      Compressibility, $\kappa_{\rm T}$   & $1.07 \times 10^{-9}\:\text{Pa}^{-1}$ \\
      Kuhn segment length, $b$            & $1.83 \:\text{nm}$ \\
      Kuhn segment molar mass             & $729.05 \:\text{g/mol}$ \\
      Estimated $A_{\rm PS}$              & $5.84 \times 10^{-20}\:\text{J}$ \\
      Estimated $A_\text{\ce{SiO2}}$      & $6.43 \times 10^{-20}\:\text{J}$ \\
      Grid edge length, $\Delta L$        & $2.5 \:\text{nm}$
   \end{tabular}
\end{table}

\section{Results and discussion}
\subsection{Local Structure}
The local density of the polymer in proximity of the surface is often employed as a measure of the 
strength of polymer-surface interactions. Radial mass density profiles and cumulative mass distributions  
are studied here for various combinations of particle radii, grafting densities, and embedding matrices 
in order to quantify how these factors affect the size and form of the corona of grafted chains, as well as
the distribution of free chains around a nanoparticle. 

\begin{figure}
   \centering
   \includegraphics[width=0.45\textwidth]{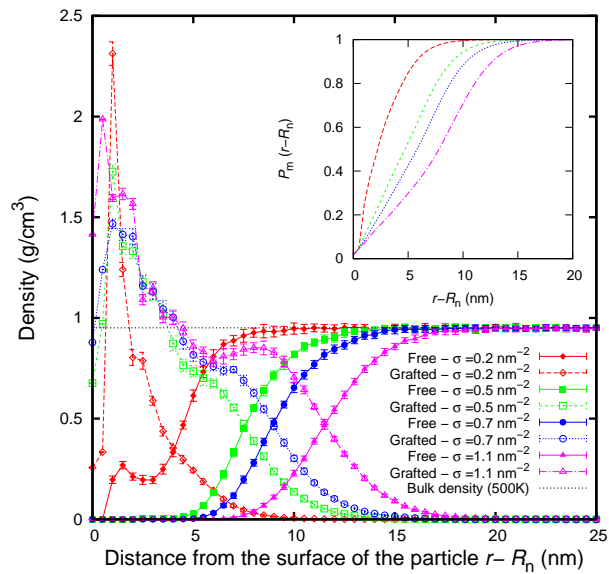} 
   \caption{Radial density profiles for free and grafted segments as a function of distance from the surface of the 
            particle. The grafting density, $\sigma$, varies from $0.2\:\text{nm}^{-2}$ to $1.1\:\text{nm}^{-2}$, 
            while the radius of the nanoparticle is $R_\text{n}=8\:\text{nm}$ and the molar mass of 
            the matrix is $M_\text{f}=100\:\text{kg/mol}$. The molar mass of grafted chains is 
            $20 \:\text{kg/mol}$. In the inset to the figure, the normalized cumulative mass 
            distribution for the grafted chains is depicted as a function of distance from the 
            surface of the particle. (Density profiles are accumulated in 0.5-nm-thick bins.)}
   \label{fig:rdf_vs_grafted_density}
\end{figure}

\paragraph{Influence of grafting density}
 \ref{fig:rdf_vs_grafted_density} depicts the effect of grafting density on the packing of free 
and grafted chains near the surface of the particle. 
Grafted chains dominate the interfacial region, thus contributing more than free chains to the total density 
calculated near the nanoparticle surface. The tendency of grafted chains to stretch away from the surface 
increases with increasing grafting density; congestion with neighboring grafted chains forces them to 
extend more. 
The density profile of the grafted segments exhibits two regimes: a parabolic decay near
the surface of the particle, and an exponential tail which extends far into the matrix.
This form is in agreement with predictions from theoretical \cite{Science_251_905},  
SCF \cite{Macromolecules_26_7214}, MD \cite{Macromolecules_44_2316,SoftMatter_7_1418} 
and off-lattice MC\cite{PhysRevE_49_3199} simulations of grafted chains in melt and solvent environments.
The presence of chemically grafted chains on the surface inhibits the approach of free chains 
to the nanoparticle, with the strength of the exclusion of free chains increasing with increasing grafting density.
Far from the nanoparticle surface we observe a smooth decay of the density profile. 
The parabolic decay is continuous for $\sigma=0.2\:\text{nm}^{-2}$, while for higher grafting densities  
the decay is manifested by two parabolic branches, which can be attributed to the concentrated polymer brush 
regime (CPB) close to the surface and the semi-dilute polymer brush regime (SDPB) away from it
\cite{Macromolecules_40_9143}.
The sum of the grafted and free chain density profiles suggests that the system is essentially incompressible, except in 
the immediate vicinity of the grafting surface. 
Accumulated mass distributions in the inset of  \ref{fig:rdf_vs_grafted_density} show the relation between 
grafting density and brush (grafted layer) propagation. 

It can been seen that higher grafting density leads 
to a more extended grafted layer, thus increasing the thickness of the polymeric brush around the nanoparticle. 
For the highest grafting density examined, $\sigma = 1.1 \:\text{nm}^{-2}$, the interfacial region has a 
thickness of at least $20 \:\text{nm}$, as evidenced from the cumulative mass distribution.
Moreover, the density profile of grafted chains density profile exhibits a second smooth maximum before its parabolic decay. 
This feature is also present in previous simulation studies. 
Atomistic simulations of 20-mer grafted chains on a 2-nm-radius particle by Ghanbari et al.\cite{Macromolecules_45_572}
exhibit this extra hump for a grafting density of $1.0\;{\rm nm}^{-2}$.
The length scale at which this feature appears depends on the segment size, but it seems to be 
rather a real effect than an artifact of the model employed in this work. The first parabolic profile (CPB regime) 
is separated from the second one (SPDB regime) by a region of thickness equal to one Kuhn length, at the end of which
this extremely shallow and smooth local maximum appears.

Borukhov and Leibler \cite{Macromolecules_35_5171} have predicted the phase diagram  for grafted 
polymer (index of polymerization $N_\text{g}$) in contact with a chemically identical polymer melt 
(index of polymerization $N_\text{f}$), by using a scaling model along with SCF calculations.
For long enough matrix chains, $N_\text{f}>N_\text{g}$, three regions exist as a function of the 
reduced grafted density $\sigma_\text{r} = \sigma/\alpha^2$ with $\alpha$ being the size of the 
monomer in their description (in our case, $\alpha \to b/2$). For $\sigma_\text{r} < N_\text{g}^{-1}$ 
the brush is in the ideal mushroom regime, for $N_\text{g}^{-1} < \sigma_\text{r} < N_\text{g}^{-1/2}$ 
the brush is ideally wet, while for $\sigma_\text{r} > N_{\rm g}^{-1/2}$ the brush is dry. 
Following the reasoning of Borukhov and Leibler, the lowest grafting density of $\sigma = 0.2\:\text{nm}^{-2}$ 
( \ref{fig:rdf_vs_grafted_density}) lies in the proximity of the ideal wet brush regime of the phase diagram.
This is evident from the fact that some free segments have penetrated the corona and lie close
to the surface of the particle, thus leading to a local maximum in the radial density distribution of free
chains, in this case located around a Kuhn length ($b=1.83$ nm)  away from the surface of the particle. 
For higher grafting densities, our brushes fall in the dry brush regime, where free chains are 
completely expelled from the surface of the particle over a distance which scales as $N_\text{g}^{1/3}\sigma^{1/3}$.
High grafting densities of end-adsorbed polymers in the presence of chemically identical matrix polymers have
been shown previously to promote autophobicity (i.e., incompatibility between the grafted chains and the matrix)
\cite{Macromolecules_29_2150,JChemPhys_115_2794,SoftMatter_7_7914}.
It is found here that increasing $\sigma$ promotes greater matrix/brush incompatibility by 
drying the polymer brush; however, the effects are much less prominent for curved surfaces than for 
planar ones \cite{Macromolecules_31_3994}, since polymers end-grafted to small spherical surfaces face 
less chain crowding moving away from the surface.

\begin{figure}
   \centering
   \includegraphics[width=0.45\textwidth]{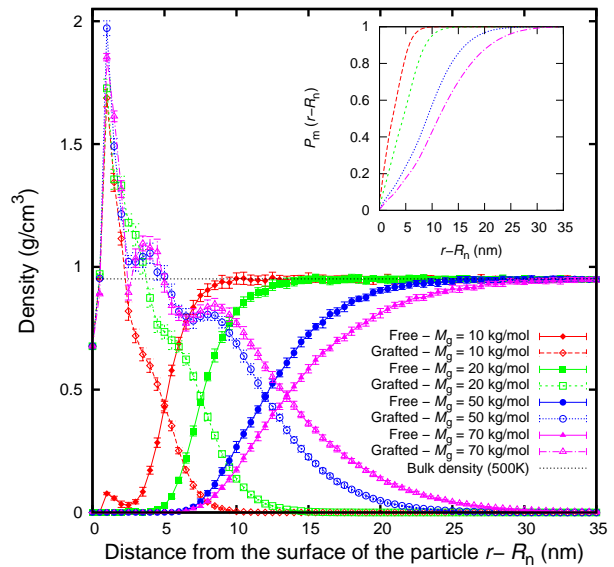} 
   \caption{Radial density profiles for grafted and free segments as a function of the grafted chain molar mass. 
            The radius of the nanoparticle is $R_\text{n} = 8\:\text{nm}$, while the molar mass of the matrix 
            is $M_\text{f}=100\:\text{kg/mol}$. The grafting density is $\sigma=0.5 \text{nm}^{-2}$.
            In the inset to the figure, the normalized cumulative mass distribution for the grafted chains 
            is depicted as a function of distance from the surface of the particle. (Density profiles 
            are accumulated in 0.5-nm-thick bins.)}
   \label{fig:rdf_vs_grafted_mass}
\end{figure}

\paragraph{Influence of grafted chain molar mass}
The influence of grafted chain molar mass on the distribution of grafted and free chains around the surface 
of the particle is examined in  \ref{fig:rdf_vs_grafted_mass}. The grafting density is kept fixed at 
$\sigma=0.5\text{nm}^{-2}$, while the molar mass of the grafted chains varies from $10\:\text{kg/mol}$
to $70\:\text{kg/mol}$. 
The profiles of the grafted chains start at the same point for $r-R_\text{n}=0$, since there their volume
density is dictated by the surface grafting density.
The grafted chain profiles exhibit almost identical density peaks, located around one Kuhn length away from the 
particle surface, since the grafting density is kept constant. 
The higher the molar mass of the grafted chains, the more the brush extends into the matrix.
For high molar mass chains ($M_\text{g}=50$ and $70 \;\text{kg/mol}$) grafted density profiles 
exhibit a second peak, attributed to the second neighbors of the segments which are permanently attached 
to the surface of the nanoparticle. Long grafted chains lead to the formation of a dry brush around the 
particle and only for $M_\text{g}=10 \:\text{kg/mol}$ do some free segments lie inside the grafted corona, 
wetting the corona.
The phase behavior of the grafted corona is in accordance with the already mentioned scaling theory of Borukhov and 
Leibler \cite{Macromolecules_35_5171}.
Nanoparticle/polymer miscibility should be generally high when the wet brush conditions are met and 
decrease under dry brush conditions.
The degree of overlap between grafted and free segments can be quantified by calculating an interpenetration parameter,
$\delta$, as this was defined by Egorov and Binder.\cite{JChemPhys_137_094901}
From our simulations, the interpenetration width extends from $7.2\;{\rm nm}$ for $10\;{\rm kg/mol}$ grafted chains, 
up to $23.5\;{\rm nm}$ for $70\;{\rm kg/mol}$ grafted chains.

\begin{figure}
   \centering
   \includegraphics[width=0.45\textwidth]{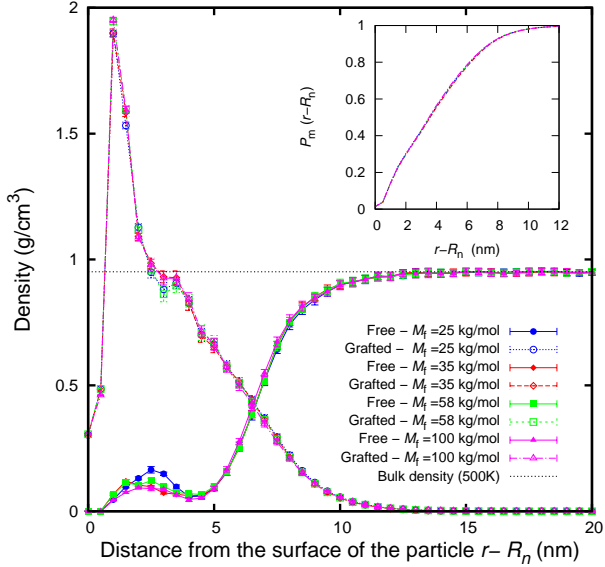} 
   \caption{Radial mass density distribution of free and grafted segments from the surface of the 
   nanoparticle, as a function of the matrix molar mass. The radius of the particle is $13 \:\text{nm}$, 
   the molar mass of the grafted chains $M_{\rm g}=25 \:\text{kg/mol}$ and the grafting density is
   $\sigma = 0.2 \text{chains/nm}^2$. 
   Matrix molar mass varies from $M_\text{f}=25 \:\text{kg/mol}$ to $M_\text{f}=100\:\text{kg/mol}$.
   In the inset to the figure, the normalized cumulative mass distribution for the grafted chains 
   is depicted as a function of the distance from the surface of the particle. (Density profiles 
   are accumulated in 0.5-nm-thick bins.)}
   \label{fig:rdf_vs_matrix_mass}
\end{figure}

\paragraph{Influence of matrix molar mass}
 \ref{fig:rdf_vs_matrix_mass} depicts the effect of matrix material on the spatial extent of grafted chains 
for identical particle radii and grafting densities. Since the molar mass of the grafted chains and
the grafting density are low enough, free chains can slightly wet the grafted corona. 
The effect is stronger for the shortest matrix chains, as expected, but remains present for all
matrix molar masses examined. 
The specifications of the systems studied are chosen as close as possible to 
the experiments of Chevigny et al. \cite{Macromolecules_43_4833}, in which it is stated that  a ``wet to dry''
conformational transition occurs. However, based on  \ref{fig:rdf_vs_matrix_mass}, a conformational
transition of the grafted chains can not be observed. The cumulative mass distributions, shown in the
inset to  \ref{fig:rdf_vs_matrix_mass}, are almost identical for all matrix molar masses. 
Since $R_\text{n}/R_\text{g} > 1$, the theory of Trombly and Ganesan \cite{JChemPhys_133_154904} 
predicts that the ``wet to dry'' transition should occur for $N_\text{f}/N_\text{g}<1$. This requirement 
would be fullfilled by studying even shorter matrix chains. Unfortunately, the coarse character of 
our methodology cannot allow us to study shorter polymeric chains. In this regard, the trend 
predicted by the theory seems to be in accordance with our simulation and qualitatively 
consistent with the experimental data. According to our single particle simulations, the brush 
profiles are almost insensitive to  changes in $N_\text{f}/N_\text{g}$ when 
$N_\text{f}>N_\text{g}$. This suggests that the experimentally observed phase behavior of these
systems may be sensitive to subtle changes in the distribution of brush and matrix segments when
two or more particles come together. Also, from the experimental point of view, a more exact 
comparison with theoretical predictions and simulations would require monodisperse 
nanoparticles and highly monodisperse polymers. 

\subsection{Scaling of grafted polymer layers}

\paragraph{Height of the grafted polymer brush}
Spatial integration of the radial mass density profiles allows for estimating the height of the grafted 
polymer brush, which is usually defined as the second moment of the segment density distribution, 
$\rho(r)$, as \cite{Macromolecules_25_2890,Macromolecules_26_7214}:
\begin{equation}
\left\langle h^2 \right\rangle^{\frac{1}{2}} = \left[\frac{\int_{R_n}^{\infty} 4 \pi r^2 dr (r-R_{\rm n})^2 \rho(r)}
{\int_{R_n}^{\infty} 4 \pi r^2 dr \rho(r)}    \right]^{\frac{1}{2}}
\label{eq_average_brush_height}
\end{equation}
with respect to the height $h \equiv r - R_\text{n}$. We therefore use this root mean square (rms) height to 
define the brush thickness. However, comparison with experimental brush heights requires a 
measurement of where the major part of the grafted material is found. To this effect, the brush height 
can also be arbitrarily defined as the radius marking the location of a spherical Gibbs dividing 
surface, in which 99\% of the grafted material is included. 
Table \ref{tab:brush_height_ratio} reports the two estimates of the brush height for all systems 
considered in this study, for different grafting densities and molar masses of the free and 
grafted chains.

\begin{table*}[t]
   \caption{Brush heights for different silica nanoparticles, free chains molar mass, 
   grafted chains molar mass and grafting densities. Also, brush thickness estimations based on 
   fitting SANS spectra by Meyer \cite{MeyerThesis} have been included.}
   \label{tab:brush_height_ratio}
   \begin{tabular}{cccccccc} \hline
   $M_{\rm f}$ & $M_{\rm g}$ & $R_\text{n}$ & $\sigma$ 
   & \multicolumn{2}{c}{$\left \langle h^2 \right \rangle^{\frac{1}{2}}\: \left(\text{nm}\right)$} 
   & \multicolumn{2}{c}{$h_{99 \%}(\text{nm})$ }\\
   $(\text{kg/mol})$ & $(\text{kg/mol})$ & $(\text{nm})$& $\left(\text{nm}^{-2}\right)$ 
   &FOMC& SANS\cite{MeyerThesis} & FOMC & SANS\cite{MeyerThesis} \\ \hline
   100 & 20 & 8 & 0.2 & 2.98 & 3.3-3.3 & 16.49 & 16.3-17.2 \\
       &    &   & 0.4 & 4.49 & 4.1-5.0 & 18.99 & 19.2-19.6 \\
       &    &   & 0.5 & 5.18 & 4.4-5.7 & 20.33 & 19.9-20.9 \\
       &    &   & 0.7 & 6.25 & 4.9-6.8 & 21.76 & 21.2-23.1 \\
       &    &   & 1.1 & 8.02 &       & 24.13 &    \\
   \hline
   100 & 10 & 8 & 0.5 & 3.05  & 2.8-3.8 & 16.22 & 16.0-17.3 \\
       & 20 &   &     & 5.26  & 4.4-5.7 & 20.19 & 19.9-20.9 \\
       & 50 &   &     & 10.30 & 7.8-9.1 & 30.87 & 27.5-28.4 \\
       & 70 &   &     & 13.08 &       & 36.85 &         \\
   \hline
   25  & 25 & 13 &0.24 & 4.80  &                         & 27.22 \\      
   35  &    &    &     & 4.74  &                         & 27.21 \\
   58  &    &    &     & 4.74  &                         & 27.22 \\
   100 &    &    &     & 4.73  &                         & 27.22 \\
   \hline  
   \end{tabular}
\end{table*}

The theory of spherical polymer brushes was pioneered by Daoud and Cotton \cite{JPhys_43_531}.
In analogy to the scaling model developed by Alexander and de Gennes for planar interfaces,
Daoud and Cotton developed a model for spherical surfaces through geometric considerations
based on starlike polymers. 
The spherical brush is divided into three regions, an inner meltlike core region, an intermediate
concentrated region (dense brush), and an outer semidilute region (swollen brush).
Daoud and Cotton predicted for star shaped polymers in the matrix a change in the scaling 
behavior as the blobs of the chains become non-ideal. 
The density profile is directly related to the average brush height, $h$, or the extension of the
corona chains. 
Neglecting the contribution of the core to the radius of the star, they found that:
\begin{equation}
h \sim N_\text{g}^{1/2} \sigma_{\rm r}^{1/4} b
\label{eq_daoud_and_cotton}
\end{equation}
Although the former relation exhibits ``ideal'' scaling with respect to the chain length dependence, the presence 
of the factor $\sigma_{\rm r}^{1/4}$ shows that the radius is in fact larger than it would be for a single 
linear chain. Thus, although we are in a regime where the chain seems to be ideal, the structure is actually 
stretched. 

\begin{figure}
   \includegraphics[width=0.45\textwidth]{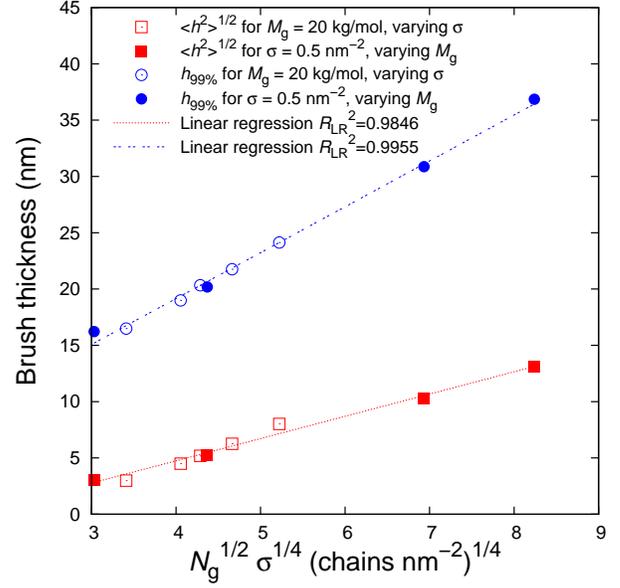}
   \caption{The calculated brush thickness (either $\left\langle h^2 \right\rangle^{\frac{1}{2}}$ or $h_\text{99\%}$)
            is plotted versus 
            the degree of polymerization of grafted chains, $N_\text{g}$, times the grafting density, $\sigma^{1/4}$. 
            Points correspond to all systems containing an 8-nm-radius particle of Table \ref{tab:brush_height_ratio}. 
            Linear behavior is predicted by the model proposed by Daoud and Cotton ( \ref{eq_daoud_and_cotton}) 
            for star shaped polymers \cite{JPhys_43_531}.} 
   \label{fig:brush_height_scaling}
\end{figure}

In  \ref{fig:brush_height_scaling} the average thickness of all analyzed systems is plotted 
versus $N_\text{g}^{1/2}\sigma^{1/4}$. $N_\text{g}$ is measured in Kuhn segments 
per chain and $\sigma$ in $\text{nm}^{-2}$. The scaling prediction of Daoud and Cotton seems
to be fullfilled for both the rms height $\left\langle h^2 \right\rangle^{\frac{1}{2}}$ and the height
containing 99\% of the brush material, $h_\text{99\%}$.
The dashed lines are linear fits, confirming the good agreement of the simulation data with 
the theoretical scaling behavior. 
The agreement seems to be better for the $h_\text{99\%}$ data points. This was expected, since 
the average brush thickness, as defined in  \ref{eq_average_brush_height}, is more sensitive
to the discretization of the model and to the post processing of the data, than the straightforward
definition of the shell in which the 99\% of the brush material can be found. 
The least squares linear regression analysis of $h_\text{99\%}$ data is more successful, 
yielding a coefficient of determination \cite{PrinciplesAndProceduresOfStatistics}, $R_\text{LR}^2$, 
higher than the corresponding for $\left\langle h^2 \right\rangle^{\frac{1}{2}}$ data points. 

\subsection{Scattering Predictions}
While previous works have focused on a single measure of the spatial distribution of polymeric chains, 
the mean brush height, examining the full distribution of polymeric segments is a more sensitive and 
critical test of the theories and simulations. This type of comparison is best performed against scattering 
patterns from Small Angle Neutron Scattering (SANS). SANS has been shown to be the most suitable technique 
to reveal the microstructure of polymers in confined environments, thanks to the specific contrast variation 
method which allowed the measurement of the form factor of a single chain through 
the use of index-matched fillers \cite{SoftMatter_5_7741,Macromolecules_43_9881}.
Neutron wavelengths are commensurate with interatomic separations, while neutron 
energies are of the same order as molecular vibrational energies. 
When used with partially deuterated polymers, SANS permits a close monitoring of macromolecular conformations 
in polymer solutions, melts, and blends. This advantage has made it a unique tool for the 
understanding of the morphology of polymer materials and of the relation between their structures and 
physical properties. Elastic scattering of neutrons measures spatial correlations between scattering centers; for polymers 
this enables the conformation to be determined.
The incoherent cross-section is isotropic and does not depend on a phase term; therefore no information can be obtained from it about 
the relative positions of the nuclei in an array. Information about the relative positions of the nuclei can be obtained if coherent 
scattering is measured. The total amplitude from the sample in the detector can be written as:
\begin{equation}
A\left(\mathbf{q}\right) = b_{\rm c} \sum_{k=1}^{\mathcal{N}} e^{-i \mathbf{q} \cdot \mathbf{r}_k} \;\;.
\end{equation}
with $b_{\rm c}$ being the mean coherent scattering length, $k$ running over all $\mathcal{N}$ nuclei of the sample
and the vector $\mathbf{q}$ which completely 
characterizes the scattering geometry: the incident and scattered beam directions and the wavelength. 
The normalized amplitude defined here is related to the intensity by:
\begin{equation}
I\left(\mathbf{q}\right) = \left\langle \left| A\left(\mathbf{q}\right) \right|^2\right\rangle 
= \left\langle \left| A(\mathbf{q}) A^{*}(\mathbf{q}) \right| \right\rangle 
\end{equation}
where the angle brackets $\left\langle \cdots \right\rangle$ indicate that the intensity observed is a
time average.

\paragraph{Single chain scattering}
To connect FOMC simulations with experimentally measured coherent scattering intensity, 
the single chain form factor can be calculated from the real space positions of polymeric segments as:
\begin{equation}
P \left(\mathbf{q}\right) = \frac{1}{(N+1)^2}\left\langle\left| \sum_{s=0}^{N} 
e^{-i \mathbf{q} \cdot \mathbf{r}_j(s)} \right|^2\right\rangle_{n}
\label{eq_form_factor_definition}
\end{equation}
where $N=N_\text{f}, n=n_\text{f}$ for the free and $N=N_\text{g}, n=n_\text{g}$ for the grafted 
chains, respectively. With $\mathbf{r}_j(s)$ we dente the $s$-th segment of the $j$-th chain.
The average is taken over all the free or grafted chains,
 across different configurations of the system. At equilibrium, the ensemble average
equals the time average which is measured experimentally.

The form factor of a freely jointed chain, which follows Rayleigh random walk statistics, can be analytically calculated by 
combining and extending the work of Chandrasekhar \cite{RevModPhys_15_1}, Daniels \cite{ProcRoySocEdinb_63_290} 
and Burchard and Kajiwara \cite{ProcRSocLondonSerA_316_185}:
\begin{align}
\label{eq_rayleigh_form_factor}
P_{\text{Rayleigh}} (q) = \frac{2}{\left(N+1\right)^2} 
& \left[  \frac{N+1}{1 - \frac{\sin(qb)}{qb}} -\frac{N+1}{2} \right . \nonumber \\
& \left .
- \frac{1 - \left(\frac{\sin(qb)}{qb} \right)^{N+1}}{\left(\frac{\sin(qb)}{qb} \right)^2}
\cdot \frac{\sin(qb)}{qb} \right]
\end{align}
where $N$ again refers either to free or grafted chains and $b$ being the Kuhn length. For $N >> 1$, chains follow Gaussian statistics
with their form factor given by the well known Debye expression:
\begin{equation}
P_{\text{Debye}}(q) = \frac{2\left(e^{-R_\text{g,0}^2 q^2} -1 +R_\text{g,0}^2 q^2 \right)}{\left(R_\text{g,0}^2 q^2\right)^2}
\label{eq_debye_form_factor}
\end{equation}
where $R_\text{g,0}^2$ is the unperturbed mean squared radius of gyration given by 
$R_\text{g,0}^2 = Nb^2/6$, where $N$ is the number of statistical (Kuhn) segments along the contour
and $b$ is the Kuhn length of the chain.
The Debye form factor is valid under the additional condition of $qb<<1$.
Since it exhibits a plateau at high $q$, it cannot capture the rise of the form factor of real polymers due to
their stiffness.
The advantage of the former equation ( \ref{eq_rayleigh_form_factor}) over the well-known Debye 
form factor ( \ref{eq_debye_form_factor}) stems from the incorporation of the finite extensibility 
of freely jointed chains, which represent real chains better than the Gaussian model at large extensions.
However, $P_{\text{Rayleigh}}(q)$  is not a well-behaved function at large $N$, rendering its algebraic 
manipulation difficult. 

\begin{figure}
   \includegraphics[width=0.45\textwidth]{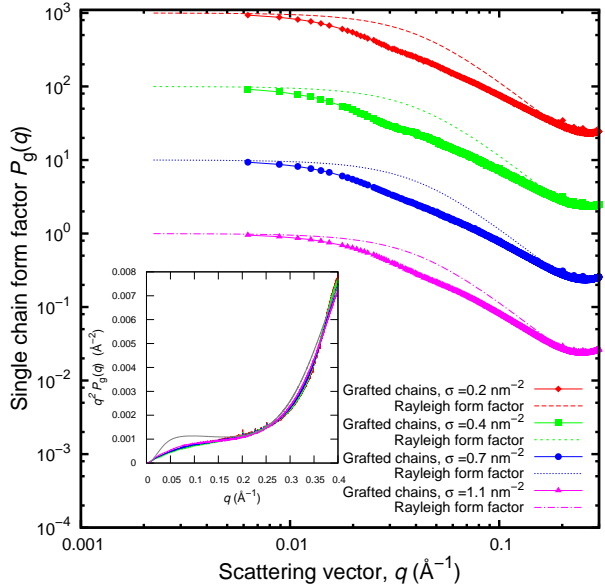}
   \caption{Form factor, $P_\text{g}(q)$, of 20 kg/mol grafted chains as a function of grafting 
            density. Systems consist of  an 8 nm-radius grafted particle immersed in 100 kg/mol 
            matrix. The Kuhn length $b$ is $1.83\:\text{nm}$ and the unperturbed radius of gyration 
            of grafted chains $R_\text{g,0}$ is $3.91\:\text{nm}$. The curves for 
            $\sigma=0.4,\:0.7,\:1.1\;\text{nm}^{-2}$ have been vertically displaced by one, two and three 
            decades respectively, for clarity.
            In the inset to the figure, the corresponding Kratky plot is presented. Dashed lines 
            display the analytically calculated Rayleigh form factor from  \ref{eq_rayleigh_form_factor}.}
   \label{fig:form_factor_vs_grafting_density}
\end{figure}

In \ref{fig:form_factor_vs_grafting_density}, the effect of grafting density on the calculated form 
factor of grafted chains is depicted. The form factor of grafted chains, calculated by 
 \ref{eq_form_factor_definition} during the course of the FOMC simulation, is compared to the 
theoretically predicted form factor from  \ref{eq_rayleigh_form_factor}. It is evident that grafted 
chains deviate from their unperturbed melt configuration. The attachment of their one end to the 
surface of an excluded volume sphere strongly affects their scaling behavior at low $q$-values. 
While at small length scales (large $q$, see Kratky plot in the inset to the figure) their conformations
are close to unperturbed, at large length scales, their confinement dictates their behavior. 
The deviation of the form factor of grafted chains from that of unperturbed freely jointed chains does
not depend monotonically on surface grafting density. The radius of the particle sets the length scale
($q\sim (2\pi)/R_{\rm n}$) where the deviation is manifested, which is common across the different systems
depicted in \ref{fig:form_factor_vs_grafting_density}.  

\begin{figure}
   \includegraphics[width=0.45\textwidth]{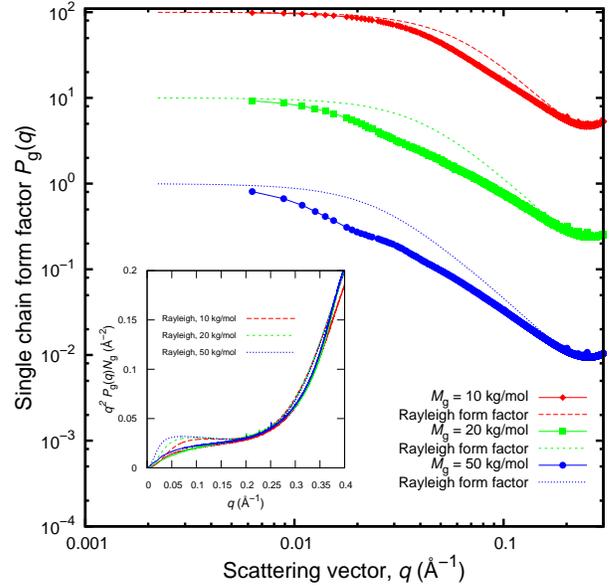}
   \caption{Form factor, $P_\text{g}(q)$, of grafted chains as a function of their molar mass. 
            Systems consist of an 8 nm-radius particle at $\sigma = 0.5 \:\text{nm}^{-2}$ surface 
            coverage immersed in 100 kg/mol matrix. The curves for $M_\text{g}=20$ and $50\;\text{kg/mol}$
            have been vertically displaced by one and two decades, respectively, for clarity.
            The unperturbed radii of gyration, $R_\text{g,0}$, are $2.77\;\text{nm (10 kg/mol)}$, 
            $3.91\;\text{nm (20 kg/mol)}$ and $6.19\;\text{nm (50 kg/mol)}$. 
            In the inset to the figure, the Kratky plot is presented, where form factors have been scaled with 
            chain length in order the plateaus to coincide for all systems.  
            Dashed lines correspond to analytically 
            calculated Rayleigh form factors from  \ref{eq_rayleigh_form_factor}.}
   \label{fig:form_factor_vs_grafted_mass}
\end{figure}

The effect of increasing the grafted chain molar mass on the calculated scattering from single grafted
chains is examined in  \ref{fig:form_factor_vs_grafted_mass}. 
For all molar masses examined, chains deviate strongly from their unperturbed configurations.
The confinement affects their behavior at low and intermediate $q$-values, while leaving large
$q$-values (small length scales) unaffected. 
The permanent link between one of their ends and the particle surface and the  excluded
volume of the particle forces the grafted chains to extend more into the bulk.
This tendency increases with increasing grafted chain molar mass; there, the deviation of 
grafted single chain scattering from the theoretical Rayleigh prediction becomes more pronounced.
Moreover, this deviation moves systematically to smaller $q$-values as $M_{\rm g}$ increases.
By careful examination one deternines that the lowest $q$-value at which this deviation is manifested is around 
$qR_\text{g}\sim 2$, with $R_\text{g}$ being the root mean squared unperturbed radius 
of gyration, $R_\text{g}=\sqrt{R_\text{g,0}^2}$. This implies that all chains deform due to their
confinement in the same way, with the actual length scale of the deformation following their 
unperturbed dimensions. A similar observation was made by Grest et al.\cite{Macromolecules_20_1376} who 
calculated from MD simulations the scattering from single chains within a star polymer. 
At higher $q$ values, the radius of the particle sets the length scale of the scattering in this region, 
causing grafted chains to deviate from their unperturbed configurations and pushing them to acquire
conformations which are closer to a shell-like structure than to an uncostrained random walk. 

\begin{figure}
   \includegraphics[width=0.45\textwidth]{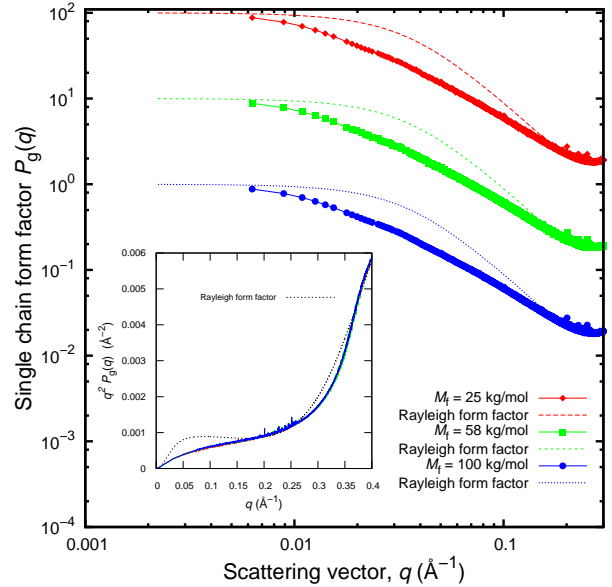}
   \caption{Form factor, $P_\text{g}(q)$, of grafted chains as a function of the surrounding
            matrix molar mass. Systems consist of an 8 nm-radius particle with grafted chains of $M_\text{g}$ = 20 kg/mol 
            at $\sigma = 0.5 \:\text{nm}^{-2}$ surface coverage immersed in matrices of $M_\text{n,f}=25$, $58$ and $100 \:\text{kg/mol}$.
            In the inset to the figure, the Kratky plot is presented. The dotted line corresponds 
            to the analytically calculated Rayleigh form factor from  \ref{eq_rayleigh_form_factor}.}
   \label{fig:single_vs_matrix_mass}
\end{figure}

In  \ref{fig:single_vs_matrix_mass} the grafted chain form factor is shown as a function of the molar
mass of the surrounding matrix. The scattering of grafted chains depends weakly on the molar mass
of the matrix in which the nanoparticles are immersed. However, it seems that the grafted chains immersed
in the matrix of molar mass 25 kg/mol deviate more at low $q$-values than in the matrices of higher molar mass. 
This implies that low molecular weight grafted chains are further extended in the presence of low molar mass free chains, 
due to their interpenetration with the grafted corona, as was seen by radial density distribution also. 

\begin{figure}
   \includegraphics[width=0.45\textwidth]{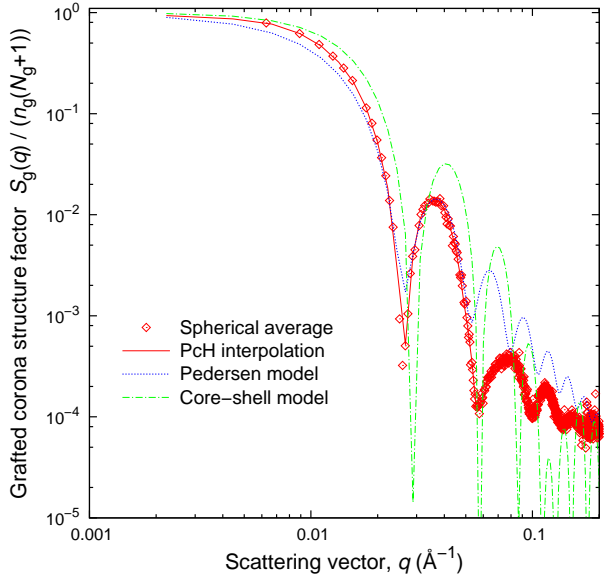}
   \caption{Scattering curve, $S_\text{g}(q)$, of a grafted corona consisting of $20\:\text{kg/mol}$
            chains, grafted onto a $8 \text{nm}$-radius particle, at $0.5\:\text{nm}^{-2}$ surface
            density, immersed in a $100\:\text{kg/mol}$ matrix. Simulation results are presented 
            along with several theoretical models.}
   \label{fig:corona_models_comparison}
\end{figure}

\paragraph{Corona scattering}
In FOMC simulations, the grafted corona structure factor can be calculated by summing the contributions 
of all grafted segments and then  analyzed by comparison to theoretical models and experimental
results, obtained through matching the silica core, in order to see only the grafted corona. 
The corona structure factor, $S_\text{g}(\mathbf{q})$, is evaluated as:
\begin{equation}
S_\text{g}\left(\mathbf{q}\right) = \frac{1}{n_\text{g}\left(N_\text{g}+1\right)} \left \langle
\left| \sum_{j=1}^{n_\text{g}} \sum_{s=0}^{N_g} e^{-i \mathbf{q} \cdot \mathbf{r}_j(s)} \right|^2
\right\rangle 
\end{equation}
where the double sum goes over all segments belonging to grafted chains. 
In  \ref{fig:corona_models_comparison} the resulting curve for a grafted corona is shown. 
The points represent the average over all $\mathbf{q}$-vectors
sharing the same norm. In order to get an impression of the scattering curve, a Piecewise cubic 
Hermite (PcH) spline interpolation scheme is used \cite{ComputerAidedGeometricDesign}. 
Along with the simulation results, the behavior of  two theoretical models is also illustrated in 
\ref{fig:corona_models_comparison}. The first one is the form factor of a spherical shell of
uniform density and thickness equal to the estimated brush thickness, $\left\langle h^2 \right\rangle^{1/2}$
\cite{PolymersAndNeutronScattering}.
The other one is a model initially developed by Pedersen and Gestenberg
\cite{Macromolecules_29_1363,JChemPhys_114_2839} for block copolymer micelles. 
On the average, the constant density shell form factor exhibits the well-known dominant $q^{-4}$ 
behavior. The periodic steps suggest a rather sharp cutoff in the segment density at radius
equal to $R_\text{n}+\left\langle h^2 \right\rangle^{1/2}$, which is a crude approximation to the real brush,
since the brush density distribution actually exhibits a smooth and continuous variation as a function of distance 
from the surface of the particle. 
The assumption that the brush is a homogeneous spherical shell of thickness $\left\langle h^2 \right\rangle^{1/2}$
shifts the position of the structure factor minimum to larger $q$-values (smaller length scales), 
compared to its actual position.

A better approximation to the corona structure factor is the model introduced by Pedersen.
This model is based on several different terms which have to be determined: the self 
correlation of the spherical core, the self correlation of the chains, the cross term between 
spheres and chains, and the cross term between different chains. 
In contrast-matching conditions for which the silica core is matched to see only the polymer 
corona, the Pedersen model is reduced to a weighted sum of the single chain form factor and 
the intra-chain cross correlation form factor:
\begin{equation}
S_{\text{Pedersen}}(\mathbf{q})=n_\text{g} \cdot \left(P(\mathbf{q}) 
+ (n_\text{g}-1) S_{\text{chain-chain}}(\mathbf{q}) \right)
\end{equation}
where $P_{\rm g}(\mathbf{q})$ is the single chain form factor of the grafted chains
and $S_\text{chain-chain} (\mathbf{q})$ is the cross term between different chains.
In our approach, the single chain form factor can be directly calculated  
from the segmental positions ( \ref{eq_form_factor_definition}) and therefore can be used 
for the evaluation of Pedersen structure factor, thus taking 
into account the stretching of the chains due to the particle. The cross term related to the interaction between 
chains which are assumed to be evenly distributed on the surface of the sphere and follow Gaussian 
statistics can be analytically expressed as \cite{Macromolecules_29_1363}:
\begin{equation}
S_{\text{chain-chain}}(q) = \left[\frac{1-e^{-qR_g}}{q R_g} \right]^2 \; 
\left[\frac{\sin{qR_n}}{qR_n} \right]^2 \;\;.
\end{equation}
Alternatively, for freely jointed chains following the Rayleigh distribution, a numerical integration of the cross-term between 
two chains obeying  \ref{eq_rayleigh_form_factor} should be carried out.
In the derivation of the Pedersen structure factor, the chains are free to penetrate into the core. 
In order to mimic the presence of the excluded volume of the nanoparticle, the starting points 
of the chains are shifted from the distance $R_\text{n}$ to $R_\text{n} + a R_\text{g}$, 
with $a$ close to unity \cite{JChemPhys_114_2839}.

\begin{figure}
   \includegraphics[width=0.45\textwidth]{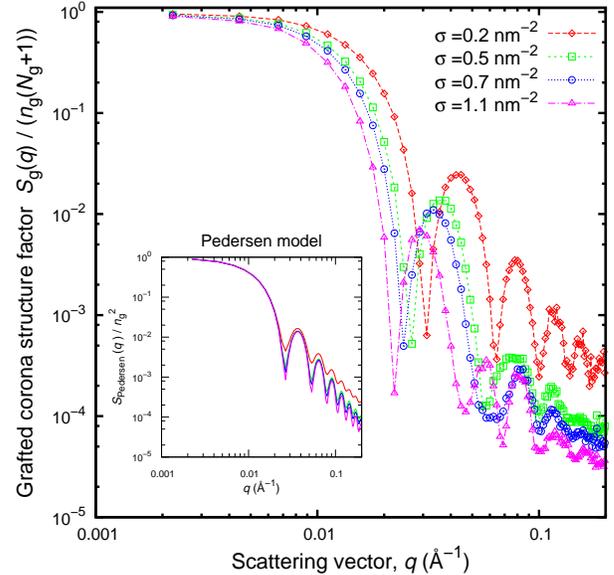}
   \caption{Scattering curves of the deuterated grafted corona ($M_\text{g} = 20\:\text{kg/mol}$) inside 
            the silica matched matrix ($M_\text{f}=100 \:\text{kg/mol}$). The radius of the particle is 
            $R_\text{n}=8 \:\text{nm}$ while the grafting density varies from $0.2\:\text{nm}$ to 
            $1.1\:\text{nm}$. }
   \label{fig:corona_vs_grafting_density}
\end{figure}

The corona scattering curves are shown in  \ref{fig:corona_vs_grafting_density} as a function of the
grafting density. All curves are normalized so as to approach unity for $q\to0$.
At the lowest $q$-values, the intensity reaches a plateau, indicating finite-size objects, which
confirms  that there is no interference between the periodic images of the particles. 
At intermediate $q$-values, an oscillation is present with its position being characteristic of the 
corona size \cite{PhysRevLett_77_95}: the thicker the corona, the smaller the value of $q$  at which the 
sharp oscillation appears. 
Here again, our results  support the fact that increasing $\sigma$ leads to more extended grafted brushes. 
In the inset of  \ref{fig:corona_vs_grafting_density} is shown the behavior of the Pedersen model with 
non-interacting chains obeying Rayleigh statistics. 
At low grafting densities, the simulated scattering curves are close to the ones predicted by the 
Pedersen model. At higher grafting densities the simulation results deviate from the model. 
The Pedersen model considers chains which are stretched due to the excluded volume of the particle, but
are not stretched due to crowding with their end-grafted partners. This assumption is 
valid for low enough grafting densities, where the dominant non-ideal contribution comes from the 
particle's excluded volume. As the neighborhood becomes more crowded, however, the Pedersen model becomes incapable
of capturing the behavior of the grafted chains, since it does not consider chain-chain interactions. 
The brush thickness of the model does not depend on the grafting density, as can be seen from 
the inset to  \ref{fig:corona_vs_grafting_density}, where the position of the oscillation is roughly 
the same for all systems.

\begin{figure}
   \includegraphics[width=0.45\textwidth]{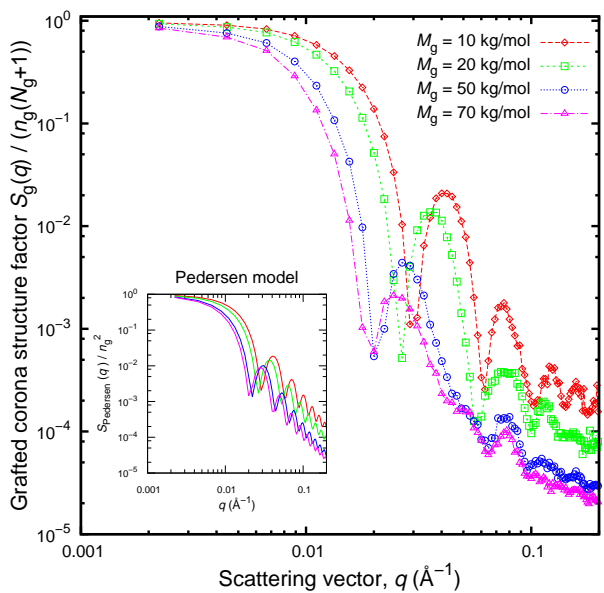}
   \caption{Scattering curves of the deuterated grafted corona inside the silica matched matrix
            ($M_\text{f}=100 \:\text{kg/mol}$). 
            The molar mass of the grafted chains varies from $10$ to $70\:\text{kg/mol}$. 
            The radius of the particle is $R_\text{n}=8 \:\text{nm}$ with grafting density of 
            $0.5 \:\text{nm}^{-2}$.}
   \label{fig:corona_vs_grafted_mass}
\end{figure}

The effect of grafted molar mass on the scattering of the corona is illustrated in 
\ref{fig:corona_vs_grafted_mass}. In analogy to the figure before, the structure factor of the
grafted corona is presented under silica-matched conditions. 
It is evident that increasing the molar mass of the grafted chains increases the height of the
polymeric brush. This is clearly manifested by the shifting of the structure factor oscillation
to smaller $q$-values. 
The molar mass dependence of the corona scattering is also predicted by the Pedersen model, 
as can be seen in the inset of  \ref{fig:corona_vs_grafted_mass}, where both the position and the
shape of the oscillation changes as a function of the chains molar mass. 
As the molar mass of the grafted chains increases, the ``terraced'' profile of $S_\text{g}(q)$ 
becomes less structured and falls off very rapidly. This is due to the high interpenetration 
of the chains as the brush becomes denser.

\begin{figure}
   \includegraphics[width=0.45\textwidth]{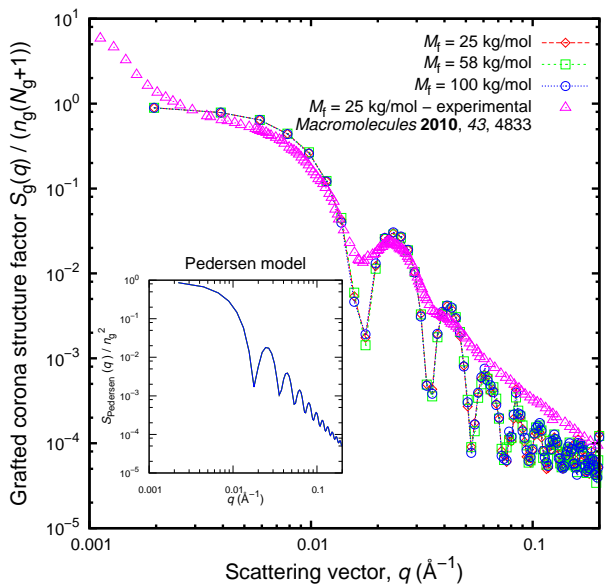}
   \caption{Scattering curves of the grafted corona ($M_\text{g} = 25\:\text{kg/mol}$) inside the silica matched matrix.
   The radius of the particle is $13\:\text{nm}$ while the molar mass of the matrix varies from $25$ to $100\:\text{kg/mol}$.
   Experimental results by Chevigny et al. for the same system are also presented \cite{Macromolecules_43_4833}.}
   \label{fig:corona_vs_matrix_mass}
\end{figure}

Finally, the influence of the matrix molar mass on the scattering of grafted chains is examined in 
\ref{fig:corona_vs_matrix_mass}. Along with the simulation results and the theoretical modeling, 
experimental results from Chevigny et al. \cite{Macromolecules_43_4833} are also presented. The 
variation of the matrix seems not to affect strongly the form of the grafted corona. 
The proposed FOMC methodology seems to capture rather well the profile of $S_\text{g}(q)$. The position
of the oscillations is the same between experimental and simulation results. However, simulation results
are more structured, with strong peak maxima and minima, which are most likely due to the inherently discrete nature of 
the FOMC model (chain mass localized in discrete Kuhn segments, rather than distributed along chain backbone).

\section{Summary and Conclusions}
Soft polymer properties are improved by inclusion of small hard inorganic particles inside the
melt matrix. The strategy of tethering polymer chain ends onto the surface of the filler 
particles, in order to render the nanoparticles miscible with their homopolymer hosts, is promising.
The structural features of such mixtures are surprisingly rich, since competing interactions 
between the nanoparticles, grafted and matrix chains engender the formation of different phases.
Overall, the phase behavior of these polymer nanocomposites may be tailored through control 
of the following parameters: grafting chain length, host chain length and grafting density, for
a given filler. In this work we have presented a methodology that is capable of treating realistic polymer 
nanocomposites at experimentally relevant length scales. The sampling, through a Monte Carlo procedure,
of a composite system obeying a simplified Hamiltonian based on polymer mean field theory, can 
give us insights  into structure at length scales on the order of hundreds of nanometers.

Our results include a number of salient points: (i) We show that by increasing the particle 
grafting density the brush of  grafted chains undergoes a phase transition from its stretched
to its collapsed form, in agreement to scaling theories\cite{Macromolecules_35_5171} and 
MD simulations\cite{SoftMatter_7_1418}. The phase transition has been attributed to the autophobic 
dewetting of the brush by the melt. 
(ii) Increasing the brush chain length leads to thicker grafted brushes, which improve the 
miscibility of nanoparticles with the homopolymer matrix, in agreement with SCF 
studies.\cite{Macromolecules_26_7214}
(iii) The density distributions around a nanoparticle seem to depend only on grafting density 
and grafted chain length, especially when the matrix chain length is equal to or longer than the
grafted chain length. ``Wet-to-dry'' conformational transition\cite{Macromolecules_43_4833} has not been
observed.
(iv) The scaling of polymeric layers grafted to nanoparticles can be well described by the model
proposed by Daoud and Cotton\cite{JPhysFrance_38_983} for star-shaped polymers. The brush thickness scales 
with the inverse second power of the grafted chain length and the inverse fourth power of the grafted density.

For the first time, a real experimental system is mapped onto a simulation framework and results which can be directly 
compared with experimental findings are provided. The coarse-grained model employed in the present work can treat 
length scales accessible by Small Angle Neutron Scattering experiments. 
Single chain and corona scattering spectra have been estimated from simulations in order 
to investigate the influence of grafted chain specifications on their conformations and overall brush thickness. 
At the single chain level, grafted chains deviate strongly from their random walk statistics, especially for high 
surface coverage and molar mass. Moreover, the scattering of the whole corona has been analyzed. The brush thickness 
increases with increasing molar mass of the grafted chains, shifting the scattering peaks to lower $q$-values, in agreement 
with the model proposed by Pedersen \cite{JChemPhys_114_2839} for the scattering of block copolymer micelles.
The corona structure factors for $13$-nm-radius grafted particles dispersed in different molar mass matrices have been 
validated against experimental findings of Chevigny et al.\cite{Macromolecules_43_4833}

The present study is limited to the case of a single nanoparticle immersed in a homogeneous polymer matrix.
As Kalb et al. \cite{SoftMatter_7_1418} stated, simulations of a single particle at infinite dilution cannot reveal 
the critical conditions that distinguish aggregation from dispersion. It is highly interesting to analyze 
the collective effects on the structure of the brush and the melt when two or more nanoparticles are present at 
experimental volume fractions. Also, the interaction and the equilibrium conformation of these systems would be of 
ultimate importance. The way of extending the present MC sampling methodology to many-particle systems is the focus of 
our current research efforts.

\begin{acknowledgement}
Part of this work was funded by the European Union through the project COMPNANOCOMP under grant 
number 295355. G.V. wants to thank Alexander S. Onassis Public Benefit Foundation for a doctoral 
scholarship. We are grateful for generous allocation of computer time on clusters of the School 
of Chemical Engineering  of the National Technical University of Athens, thoroughly 
maintained by Prof. Andreas Boudouvis and his collaborators.
Last but not least, fruitful discussions with Dr. Kostas Ch. Daoulas from Max Planck Institute
for Polymer Research, Dr. Mathias Meyer, Dr. Wim Pyckhout-Hintzen and 
Prof. Dr. Dieter Richter from Forschungszentrum J\"ulich are gratefully acknowledged. 
\end{acknowledgement}


\bibliography{manuscript}

\end{document}